\documentclass[prb,twocolumn]{revtex4}
\usepackage[dvips]{graphicx}
\usepackage{latexsym,amsmath,amssymb,bm,euscript}
\usepackage[dvips]{color}
\input{epsf}

\def\etal{~\textit{et~al.}} 
\def\ra{\rangle} 
\def\la{\langle} 
\def\up{\uparrow}
\def\dn{\downarrow}

\begin{document}

\title{Studies of non-magnetic impurities in the spin-1/2 Kagome
Antiferromagnet}

\author{Karol Gregor and Olexei I. Motrunich}
\affiliation{Department of Physics, California Institute of Technology,
Pasadena, CA 91125}

\date{\today}

\begin{abstract}
Motivated by recent experiments on ZnCu$_3$(OH)$_6$Cl$_2$, we study
the inhomogeneous Knight shifts (local susceptibilities) of spin 1/2
Kagome antiferromagnet in the presence of nonmagnetic impurities.
Using high temperature series expansion, we calculate the local
susceptibility and its histogram down to about $T=0.4J$.
At low temperatures, we explore a Dirac spin liquid proposal and
calculate the local susceptibility in the mean field and beyond
mean field using Gutzwiller projection, finding the overall picture
to be consistent with the NMR experiments.
\end{abstract}

\maketitle


\section{Introduction}
\label{sec:intro}

Spin liquids are some of the most interesting spin phases that we know
to exist theoretically.  However experimentally they are hard to achieve
because of the competition with magnetically ordered or spin-Peierls
phases.  Frustration can suppress the order, and one of the promising
systems where strong frustration occurs with natural interactions is the
nearest neighbor spin 1/2 antiferromagnet on Kagome lattice.

Herbertsmithite ZnCu$_3$(OH)$_6$Cl$_2$ contains Kagome layers of
spin-1/2 magnetic Cu$^{2+}$ moments and has emerged as one such
candidate in the recent years.
\cite{Helton06, Ofer06, Imai07, Mendels07, Vries07, Bert07, Olariu07, Lee07}
Despite the large Curie-Weiss temperature of roughly $300$K,
experiments observe no magnetic order down to $50$mK and no sign of spin
freezing or spin-Peierls transition.
Various measurements also suggest that there is no gap to excitations,
but the nature of the phase remains elusive.

From the beginning,\cite{Helton06, Ofer06} it has been appreciated
that there is significant amount of disorder in the
ZnCu$_3$(OH)$_6$Cl$_2$.
Even though impurities do not seem to produce a glassy behavior,
they can mask the intrinsic thermodynamic properties of the system
such as the spin susceptibility and specific heat.
The type of disorder has been clarified to some degree in recent
experiments,\cite{Vries07, Bert07, Olariu07} which suggest that
5 to 10\% of the Cu that would ideally be in the Kagome layers
interchange their position with the nonmagnetic Zn between the
layers.  The picture that is suggested is that the misplaced Cu behave
as weakly coupled spins giving rise to the Curie tail and anomalous
low-temperature specific heat in the absence of the field,
while the misplaced Zn effectively create spin vacancies in the
Kagome layers.

There have been many theoretical studies of the clean spin-1/2 Kagome
system.\cite{Elser, Chalker92, Leung93, Lecheminant97, Waldtmann98, Sindzingre00, Singh92, Elstner94, Misguich05, Rigol07, Misguich07, Sachdev91, Wang06, Marston91, KunYang93, Hastings01, Ran06, Ran07, Ma08, AVL, Singh07}
Exact diagonalization studies\cite{Elser, Chalker92, Leung93, Lecheminant97, Waldtmann98, Sindzingre00}
indicate no magnetic or spin-Peierls order.
Spin wave and series expansion studies\cite{Elser, Singh92}
near several ordered states also indicate that the magnetic orders
are not stable (but a recent work\cite{Singh07} suggests that a
valence bond solid may be stable).
High temperature expansion studies \cite{Elstner94, Misguich05}
for susceptibility of the Heisenberg model provide good quantitative
comparison with the experiments and recent works
\cite{Rigol07, Misguich07} account for possible Curie impurity
spin contribution or possible Dzyaloshinsky-Moriya interactions
to better match the measurements in the Herbertsmithite.

There are a number of theoretical proposals for spin liquids in the
spin-1/2 Kagome.\cite{Sachdev91, Wang06, Marston91, KunYang93, Hastings01, Ran06, Ran07, AVL, Ma08}
Of interest to the present work are gapless spin liquids.
One is the Algebraic (Dirac) Spin Liquid (ASL) of
Hastings\cite{Hastings01} and
Ran\etal\cite{Ran06} whose trial energy is very close to the ground state
energy obtained from the exact diagonalization.\cite{Ran06}
Recently, Ma and Marston\cite{Ma08} suggested that a different spin
liquid with a spinon Fermi surface may be stabilized in the
presence of ferromagnetic second-neighbor interactions.
Coming from a very different perspective, Ryu\etal\cite{AVL} suggested
an Algebraic (Dirac) Vortex Liquid, which bears some resemblance
to the ASL in its gapless nature but is a distinct phase.

Impurities modify the properties of the ground state, but at the same
time they can be used as a probe of the system, which is our pursuit
here.  In the NMR context, impurities allow the external magnetic field,
which would be strictly $q=0$ probe in a clean system, to couple to
other potentially more important wavevectors.
Despite a large body of work on defects in correlated systems
(see Refs.~\onlinecite{Alloul07, Kolezhuk06} and references therein),
there have been only few studies of impurities in frustrated
magnets.\cite{Normand02, Dommange03, Kolezhuk06, Martins07}
Exact diagonalization study by Dommange\etal\cite{Dommange03} provides
important hints on the behavior of spins near a vacancy in the
Kagome antiferromagnet.

Our work is motivated by the recent $^{17}$O NMR measurements of
Olariu\etal\cite{Olariu07}, which shows direct access to the local
spin susceptibility.
Each oxygen atom feels the magnetic moments of two copper atoms it is
connected to, see pictures in Refs.~\onlinecite{Imai07, Olariu07}.
Because of the nonmagnetic impurities, some oxygens feel two and some
only one spin.  The two peaks that are observed in the NMR spectra
at high temperatures are explained as coming from these two types of
oxygens, neglecting variations in local susceptibility.
As the temperature is lowered the peaks first move towards larger
susceptibilities and then back to the smaller ones broadening at the
same time.  At the lowest temperatures only one peak is visible
which saturates at small but positive susceptibility value.
A strong inhomogeneous broadening was also observed in $^{35}$Cl
NMR at low temperatures,\cite{Imai07} but with no clear
features that would allow spatial resolution.

In this paper we calculate local susceptibility of Kagome lattice
in the presence of nonmagnetic impurities, which are treated as
missing spins, by two different methods.

In the first part we use 12-th order high temperature series expansion
for the nearest neighbor antiferromagnetic Heisenberg model.
We calculate local susceptibility near a single impurity
Fig.~\ref{fig_local_susc_highT} and for a lattice with 5\% of
impurities Figs.~\ref{fig_HT_hist_susc}, \ref{fig_HT_hist_susc_pair}.
We obtain quantitatively accurate results down to about $T \approx 0.4J$.
These can be directly used to test whether the Heisenberg model with
spin vacancies applies to the ZnCu$_3$(OH)$_6$Cl$_2$.  For this range of
temperatures, the local susceptibility goes to the bulk value already
at a few lattice spacings from an impurity.
On the other hand, the local susceptibility $\chi_1$ of the spin
next to the impurity is larger than the bulk $\chi_{\rm bulk}$
by as much as 15-20\%.  This can be understood by realizing that the
Cu near the impurity has only three neighbors instead of four, which
directly affects the local Curies-Weiss temperature giving 14\% change
in $\chi$ at this temperature.  The series expansion study develops
this observation systematically to significantly lower temperatures
and for all sites.
On the other hand, we find that $\chi_1$ starts to decrease around
$T \approx 0.6 J$ and drops below the bulk susceptibility at
$T \approx 0.4 J$, well before the bulk susceptibility starts to
decrease at about $J/6$.\cite{Elstner94, Misguich07}
This is consistent with the ground state picture found in the exact
diagonalization study,\cite{Dommange03} where a pair of spins that are
both next to impurity forms an almost perfect singlet, so these spins
become effectively non-responsive.

One lesson we learn is that even at elevated temperatures one cannot
assume the same susceptibility on all Kagome sites.
When connecting with the experiments, we suggest to consider at least
two different susceptibilities: $\chi_1$ for the immediate neighbors
of impurities and $\chi_{\rm bulk}$ for the rest of the Cu sites.
In the $^{17}$O NMR experiments,\cite{Olariu07} we then suggest four
main groups of the oxygen Knight shifts as shown in
Fig.~\ref{fig_HT_hist_susc_pair}.  We hope that our finding and improved
experimental resolution can be used to further test the applicability
of the Heisenberg model to the Herbertsmithite.

In the low temperature study, we attempt to understand the
susceptibility histograms in the spin liquid framework, taking
up the ASL of Hastings\cite{Hastings01} and Ran\etal\cite{Ran06}
with Dirac spectrum for the spinons.
First we perform the mean field analysis treating the spinons
as free fermions hopping on Kagome lattice with impurity sites removed.
Again we calculate the susceptibility near a single impurity and in the
presence of 5\% of impurities.  We should say from the outset that
such a simplified treatment of the spinons is likely inadequate
near impurities and one should allow modifying both spinon hopping
amplitudes and fluxes near impurities.
For example, the exact diagonalization study\cite{Dommange03} finds
strong singlet correlations on bonds next to an impurity.
Furthermore, thinking about possible more accurate self-consistent
treatments suggests that the spinon hopping should also be strengthened
on these bonds.  Such modifications will certainly have significant
effect on the calculated susceptibility of the spins nearest to the
impurity.
However we argue that there are robust features in the susceptibility
histograms that are insensitive to the details of the treatment
near the impurities.  The vacancies are a source of disorder which
will induce a finite density of states at the Fermi level with
a characteristic spatial distribution.
The bulk susceptibility associated with the sites that are further
from impurities decreases as the temperature is lowered,
just as it would for the pure system where it goes all the way to
zero because of the vanishing density of states at the Fermi energy
in the Dirac spectrum.
However the impurities cause finite density of states and as more and
more sites start to feel the impurities as we lower the temperature,
which all sites eventually do since there is a finite density of
impurities,
the movement of the histogram peak to zero stops as is seen in
Fig. \ref{fig_FF_susc_pair}.
There, the detailed features seen at high temperatures depend on the
treatment of impurities and should not be taken quantitatively;
on the other hand, the overall picture of the disorder-induced broadening
of the vanishing pure Dirac susceptibility into an asymmetric histogram
should be more robust.

We also test the approach beyond the mean field by studying the
magnetization distribution in the trial spin wavefunctions obtained
by Gutzwiller projection.  This approach can reproduce nontrivial
results for the local susceptibilities near impurities in
one dimension.\cite{Eggert95}
In the Kagome system with impurities, the main effect that we find is
the increase in local susceptibility variation by about a factor of two,
while the qualitative predictions of the mean field remain unchanged.
In particular, we find that the local susceptibility can be also
negative, deviating significantly from the positive bulk value.
This can happen because there are strong antiferromagnetic correlations
in the system.

From the above discussion, it is clear that the local susceptibility
variation due to impurities is significant both at high and low
temperatures and its experimental observation can reveal a wealth
of information about the spin state.
At high temperatures, it is possible to track several features
associated with different locations relative to impurities.
In particular, the susceptibility at the oxygen sites next to
impurity should be $\chi_1$ which is different from the half
of $2\chi_{\rm bulk}$ seen by the oxygens in the bulk.
If one could improve NMR line resolution, more peaks should
be observable at high temperatures.
On the other hand, the peaks broaden as the temperature is lowered
and particularly as the system enters the correlated paramagnet
regime.  Our spin liquid study suggests that there remain no
features in the susceptibility histogram except for the overall peak
associated with the bulk sites, which first moves towards zero as in
the pure Dirac spin liquid and then stops because of the
disorder-generated nonzero density of states at the Fermi level.
This is reminiscent of what is observed in the $^{17}$O NMR
experiments, even if not in every detail; it is also qualitatively
different from the spin liquid with a Fermi surface, where there
is a finite density of states in the pure system and the impurities
would only broaden the histogram both ways without changing
significantly the bulk peak location.


\section{High Temperature Series Expansion}
\label{sec:series}

We consider spin $1/2$ nearest neighbor anti-ferromagnetic Heisenberg
model on Kagome lattice.  Local spin susceptibility at site $i$ is
given by
\begin{equation}
\chi_{\rm loc}(i) =
\frac{(g\mu_B)^2 \la S_i^z S_{\rm tot}^z \ra}{k_B T} ~,
\end{equation}
where $S_{\rm tot}^z = \sum_j S_j^z$.  We calculate $\chi$'s in the
high temperature series expansions in the presence of a single
nonmagnetic impurity, treated as a missing site (vacancy).  We also study
the Kagome system in the presence of a finite density of nonmagnetic
impurities, which we chose to be 5\% as motivated by
experiments\cite{Vries07, Bert07, Olariu07} in ZnCu$_3$(OH)$_6$Cl$_2$.

The expansion is performed to the 12-th order in $J$ (or 13-th order
in $1/T$) using the linked cluster expansion.\cite{OitmaaBook}
The outline of the procedure is as follows.  We generate all abstract
graphs up to desired size.  Then we generate all subgraphs of these
graphs, keeping track of the location of each subgraph in the graph.
We calculate the local susceptibility for each graph at each point of
the graph.  Then we subtract all the subgraphs of each graph as needed
in the linked cluster expansion to get the contribution this graph would
give when embedded into lattice.  The local susceptibility on any
lattice can be calculated by creating all possible embeddings of all
graphs and adding their contributions at every site.
In this general formulation, the lattice does not need to be regular;
in particular, this procedure applies to the Kagome lattice with
impurities.  In the practical implementation, we use the symmetries of
the underlying pure Kagome lattice for efficiency and discard the
embeddings that contain an impurity. In this way, we obtain exact $1/T$
series for the system with impurities.

One more remark about the method.  In the derivation of the linked
cluster expansion for the local susceptibility, one needs to calculate
expressions of the form ${\rm Tr}(H^a S_i^z H^b S_{\rm tot}^z H^c)$
where $a,b,c$ are integers.
Since the total spin is conserved, $[S_{\rm tot}^z, H]=0$, and
because of the cyclic property of the trace, this can always be written
as ${\rm Tr}(H^{a+b+c}  S_i^z S_{\rm tot}^z)$.  Thus every diagram can be
evaluated using this expression.

After obtaining the series, we extend it beyond the radius of
convergence using the method of Pade approximants.  We use
[5,6], [5,7], [6,6], [6,7] and expand in variable $1/(T+\alpha)$ where
$\alpha$ is usually 0.08 as in Ref.~\onlinecite{Elstner93}.
Depending on $\alpha$ one might get a pole in the expression,
and hence divergence in susceptibility even at relatively large
temperature.  This usually happens say in
one of the approximants while the others still overlap. At low enough
temperature they start diverging and we take that as a point where
the approximation stops being valid.  Different values of $\alpha$ are
tried, and sometimes it is possible to tune to a value where all the
curves overlap completely to a much lower temperature, but that is a
pathology, probably indicating that the polynomials are all the same.
For other values of $\alpha$ the curves usually start to diverge from
each other at around the same temperature.

\subsection{Single Impurity}
\begin{figure}[h]
\epsfxsize=3.5in \centerline{\epsfbox{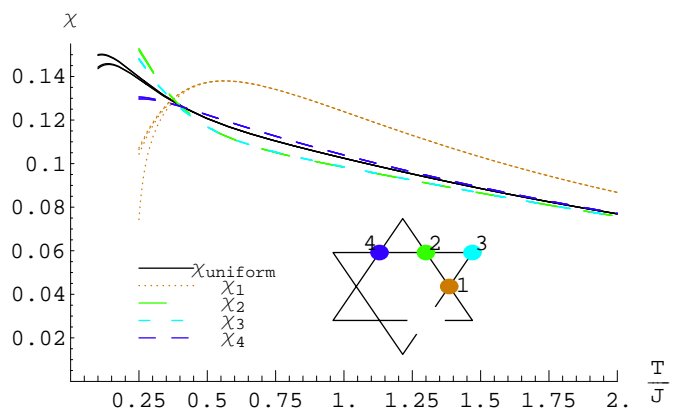}}
\caption{[Color Online]
Susceptibility of the pure Kagome system (far from any impurities)
and local susceptibilities at the first four inequivalent neighbors
of a nonmagnetic impurity.
For each $\chi_i$, there are actually four curves of the same
color corresponding to the Pade approximants [5,6], [5,7], [6,6], [6,7];
the plotting is cut where these curves start to diverge from each other.
The $\chi_2$ and $\chi_3$ happen to overlap over the entire temperature
range shown.  $\chi$ is in units $(g\mu_B)^2/J$.
}
\label{fig_local_susc_highT}.
\end{figure}

The series coefficients of the pure Kagome lattice susceptibility
and of the local susceptibilities of the first four nearest neighbors
near a single impurity are in Table~\ref{tab_local_susc}.
The corresponding susceptibilities are plotted in
Figure~\ref{fig_local_susc_highT}.

\begin{table*}
\caption{\label{tab_local_susc} Series coefficients $a_n$ of
$\chi = \frac{g^2 \mu_B^2}{T} \sum_{n=0} \frac{a_n}{4^{n+1} (n+1)\text{!}} (\frac{J}{T})^n$
for susceptibility $\chi_{\rm uniform}$ of the pure Kagome lattice
and for the four closest neighbors $\chi_i$ of nonmagnetic impurity as
indicated in Figure \ref{fig_local_susc_highT}.
The coefficients for the uniform susceptibility agree with
Ref.~\onlinecite{Elstner94} except for $n=6$, which is probably a
typo there.}
\begin{tabular}{|r|r|r|r|r|r|}
\hline n & $\chi_{\rm uniform}$ & $\chi_1$ & $\chi_2$ & $\chi_3$ & $\chi_4$ \\
\hline 0 & 1 & 1 & 1 & 1 & 1 \\
\hline 1 & -8 & -6 & -8 & -8 & -8 \\
\hline 2 & 48 & 30 & 42 & 42 & 48 \\
\hline 3 & -96 & -120 & -8 & -8 & -48 \\
\hline 4 & -320 & 680 & -360 & -240 & -1360 \\
\hline 5 & -38784 & -14592 & -45264 & -47424 & -40512 \\
\hline 6 & 677824 & 158144 & 540624 & 465248 & 995680 \\
\hline 7 & 14176256 & 2843520 & 16451456 & 18525184 & 15092736 \\
\hline 8 & -429202944 & -57813120 & -312132096 & -249716736 & -636643584 \\
\hline 9 & -11927946240 & -2620491520 & -12841824000 & -15158219520 & -12108702720 \\
\hline 10 & 501186768896 & 64507410176 & 346403354880 & 271966979328 & 700664177664 \\
\hline 11 & 13960931721216 & 3019001330688 & 14456058461184 & 17966636077056 & 13440259983360 \\
\hline 12 & -841802086780928 & -98970319958016 & -558124102202368 & -437482821907456 & -1111015315910656 \\
\hline
\end{tabular}
\end{table*}

From Fig.~\ref{fig_local_susc_highT}, we see that for the temperatures
$T \gtrsim 0.3 J$, only the first few neighbors -- most visibly the
$\chi_1$ right next to the impurity -- deviate significantly from the
susceptibility of the clean system.
The $\chi_1$ is about 15-20\% larger than $\chi_{\rm uniform}$ over a
wide temperature range $0.5 J \lesssim T \lesssim 2 J$.
As the temperature is lowered further, the $\chi_1$ starts to decrease
while the $\chi_{\rm uniform}$ still grows, and we see $\chi_1$
dropping below $\chi_{\rm uniform}$.
This is reminiscent of the picture in the exact diagonalization
study in Ref.~\onlinecite{Dommange03}, in which a pair of spins in
the same triangle as the impurity (there are two such triangles)
forms a singlet at low temperatures, and therefore the susceptibility
of these spins is strongly reduced.

\subsection{5\% Density of Impurities}
\begin{figure}[h]
\epsfxsize=3.5in \centerline{\epsfbox{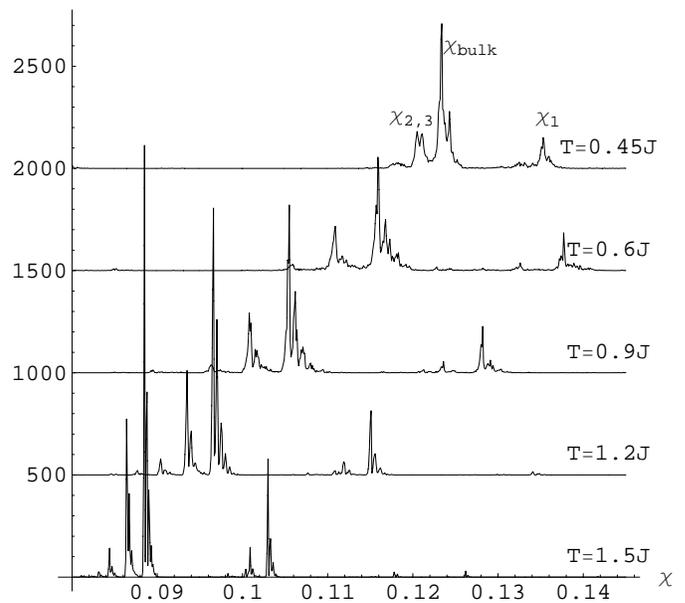}}
\caption{Histogram of local susceptibilities at the sites of Kagome
lattice in the presence of 5\% of nonmagnetic impurities.
Pade approximant [6,6] was used to analyze exact series at each site
of a large but finite $60 \times 60 \times 3$ sample with randomly
placed impurities.  $\chi$ is in units $(g\mu_B)^2/J$.}
\label{fig_HT_hist_susc}
\end{figure}

\begin{figure}[h]
\epsfxsize=3.5in \centerline{\epsfbox{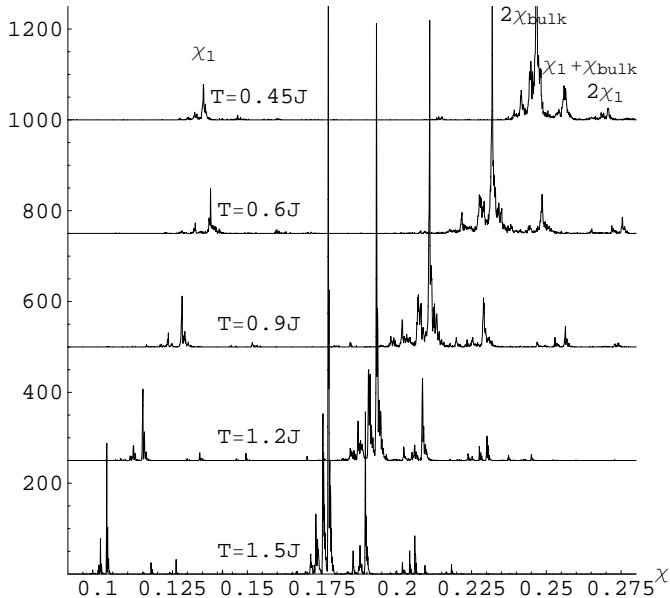}}
\caption{Histogram of the susceptibilities at oxygen sites in the
presence of 5\% of nonmagnetic impurities.
The system is the same as in Fig.~\ref{fig_HT_hist_susc} and we
used the same Pade approximant [6,6].}
\label{fig_HT_hist_susc_pair}
\end{figure}

We take a $60 \times 60 \times 3$ Kagome lattice with periodic
boundary conditions and 540 randomly placed impurities (corresponding
to 5\% density).
By considering all possible abstract graph embeddings in the
finite system, we calculate exact $1/T$ series for the local
susceptibilities in the sample.  Analyzing these as
before, we obtain reliable quantitative results down to about
$T = 0.5 J$.
The local susceptibility histograms for several temperatures
are shown in Figure~\ref{fig_HT_hist_susc}.
We also calculate the susceptibilities at the oxygen atom sites
that the experiments measure; the corresponding histograms are displayed
in Fig.~\ref{fig_HT_hist_susc_pair}.

The sample is sufficiently large that the obtained histograms are
accurate representations of an infinite system; in particular,
all features seen in Fig.~\ref{fig_HT_hist_susc} are real.
The peaks are not delta functions because all sites feel the impurities,
and the different peaks correspond to different locations relative to one
or several impurities.
We estimated the errors by dividing the sample into ten pieces,
getting the histogram for each, and calculating deviations from the
average.  The resulting relative error on the $y$ axis in
Fig.~\ref{fig_HT_hist_susc} is roughly 5\%.
There is also a systematic error along the $x$ axis coming
from the truncated series and the Pade approximants, but from the
single impurity study we believe the results are reliable for the
temperatures shown.

To the first approximation, we can say that the are two major groups
of sites: one is the sites next to impurities (susceptibilities in the
histogram near $\chi_1$), and the other is the rest of the sites
(susceptibilities near $\chi_{\rm bulk}$).
In the second group, we can also clearly see a subgroup formed by the
sites that are second neighbors to some impurity and have
susceptibilities near $\chi_2, \chi_3$.

Turning to the $^{17}$O NMR experiments, each oxygen can be
associated with a bond $<ij>$ of the Kagome lattice and feels only
these two sites $i$ and $j$.  The appropriate histogram of such
$\chi_i + \chi_j$ is in Figure~\ref{fig_HT_hist_susc_pair}.

In the first approximation, we see four major groups of peaks which
can be roughly identified as follows.  Imagine that only the
susceptibility of the sites next to the impurities is modified
and is $\chi_1$, while the rest of the Cu sites have $\chi_{\rm bulk}$
(in particular we join the $\chi_2$ and $\chi_3$ with the
$\chi_{\rm bulk}$, even though they are still separately visible in
Fig.~\ref{fig_HT_hist_susc}).
The leftmost peak in Fig.~\ref{fig_HT_hist_susc_pair} is located near
$\chi_1$ and comes from the oxygens next to impurity that feel only
one Cu site, and so the value is roughly half compared to all other
oxygens.
The dominant peak is near $2\chi_{\rm bulk}$ and comes from oxygens that
are in the bulk - neither near impurity, nor near Cu next to impurity.
The rightmost peak is located around $2\chi_1$ and comes from oxygens
that are interacting with two Cu sites that are next to impurity.
Finally, the group $\chi_1 + \chi_{\rm bulk}$ comes from oxygens that are
interacting with one site near and one not near impurity.
Again, the finer features and the finite widths of the peaks originate
from the variation of the susceptibility with the distance from
various impurities a given site feels.
The volume fractions of the four groups of peaks in
Fig.~\ref{fig_HT_hist_susc_pair} say at $T=0.6J$, are, from the left,
roughly 8\%, 70\%, 15\%, 5\%.
If we calculate how many oxygens of each type other than
$2\chi_{\rm bulk}$ are near a single impurity and then multiply it by
the density of impurities and divide by the ratio of the number of
oxygens to the number of Kagome sites (two), we would get the following
percentages 10\%, 65\%, 20\%, 5\%;
these are slightly different from the exact numbers because some sites
are close to more than one impurity.

Thinking about the experiments, the NMR lines are too broad
to resolve the groups $2\chi_{\rm bulk}$, $\chi_1 + \chi_{\rm bulk}$, and
$2\chi_1$, but just enough to separate these from the group $\chi_1$.
We hope that our study will stimulate further NMR experiments with
more site resolution.


\section{Dirac spin liquid: Mean field with impurities}
In this section we explore the Kagome system with non-magnetic
impurities at low temperatures within a particular spin liquid
proposal of Hastings\cite{Hastings01} and Ran\etal\cite{Ran06}.
In the clean system, one has fermionic spinons hopping on the Kagome
lattice in the presence of $\pi$ flux through every hexagon.
Ran\etal\cite{Ran06} showed that the corresponding spin wavefunction
obtained by Gutzwiller projection has very good variational
energy.\cite{Ran06}
Again we consider the cases of a single impurity and 5\% density of
impurities.  We take a crude model of the spin liquid in the presence
of impurities by simply removing the vacancy sites (and the links)
in the spinon hopping problem.
We will argue that while this is not fully adequate,
particularly for detailed features near impurities,
the results for the overall behavior of the bulk of the local
susceptibility histograms are robust.
We will further discuss possible improvements of our approximation
in Sec.~\ref{sec:improvements}.

In the mean field the fermions are free.  The spectrum of the clean
system has a flat band at the lowest energy containing $1/3$ of states
with gap to the next band.  The system has Dirac points at fillings
$1/2$, $2/3$ and $5/6$.  In our approach, adding impurities does not
destroy the flat band (which can be expressed in terms of localized
states), but also produces localized states in the gap.
For a finite concentration of impurities, which is our main interest,
the Dirac points are filled with finite density of states (DOS).
We consider the system at half filling which corresponds to spin $1/2$.
The pure system has Dirac nodes and a linearly vanishing density
of states at the Fermi level, while the impurities induce a finite
density of states, and one of the main questions we address is how
it is distributed locally.

If $\{ \psi_n(i) \}$ is the set of single-particle wavefunctions,
it is easy to show that the local susceptibility at temperature $T$
is given by
\begin{eqnarray}
\chi_{\rm loc}(i) &=& \frac{(g\mu_B)^2}{2 T}
\sum_n |\psi_n(i)|^2 f(\epsilon_n) (1-f(\epsilon_n))
\label{eq_chi_freeF}
\end{eqnarray}
where $f(\epsilon) = 1/(e^{(\epsilon-\mu)/T} + 1)$ is the Fermi
distribution.
For a given configuration of impurities, we obtain the single-particle
orbitals by numerical diagonalization and use this formula to calculate
the local susceptibility.
In the $T \to 0$ limit, the local susceptibility is simply proportional
to the local density of states at the Fermi level,
$\nu_{\rm loc}(i) = \sum_n |\psi_n(i)|^2 \delta(\epsilon_n - \mu)$.

Let us briefly describe the localization properties of the
single-particle states in our system with 5\% density of impurities.
We expect that all states are eventually localized, since this
is a time-reversal invariant Anderson localization problem
in two dimensions with no special symmetries.
The localization is weaker in the spectral regions where the density
of states is high and is stronger where the DOS is low, e.g., in the
vicinity of the Dirac nodes of the pure spectrum.
However, even for the latter as happens around half-filling,
the localization lengths are still very large.
For example, we calculate the participation ratios of the states near
half-filling and find that the corresponding volume fraction of
participating sites decreases only from about 14\% in a
$30 \times 30 \times 3$ system to about 10\% in a
$60 \times 60 \times 3$ system, indicating that these states are
still extended over our system sizes up to $60 \times 60 \times 3$.
On the other hand, such long distance localization physics does not
affect our measurements of local properties at finite temperatures
presented here.  For example, even a single disorder sample in a
system as small as $10 \times 10 \times 3$ produces the susceptibility
histogram which is essentially the same as from much larger samples.

Some explanations are in order before we show the results for
the local susceptibilities.
Throughout, we keep the spinon hopping $t$ fixed and vary the
temperature.
All presented susceptibilities are in units of $(g\mu_B)^2/t$.
In a more realistic calculation, the spinon hopping would need
to be found self-consistently for a given Heisenberg exchange $J$
and temperature $T$.
In such treatments of pure systems, one typically finds that the
self-consistent $t$ vanishes above some temperature of order $J$
(e.g., in the renormalized mean field scheme this temperature is
$0.75 J$).
There is no sense of speaking about spinons above this temperature,
and the system is more appropriately described in terms of
weakly correlated individual spins.  On the other hand, $t$ becomes
nonzero at lower temperatures, suggesting that the system enters the
correlated paramagnet regime.  Below the onset temperature, the
self-consistent $t$ quickly reaches the values similar to the
zero-temperature limit.  It is this regime where
$t(T) \simeq t(T=0)$ that we are describing.

We can estimate the spinon hopping amplitude from self-consistent
mean field conditions.  For example, in the so-called renormalized
mean field scheme, this reads
\begin{equation}
t_{ij} = \frac{3 g_s J_{ij}}{8}
\la f_{j\sigma}^\dagger f_{i\sigma} \ra ~,
\label{RMF}
\end{equation}
where $g_s$ is some renormalization factor (which can be found,
e.g., by comparing measurements before and after the projection).
In the case with $\pi$ flux through each hexagon and at $T=0$,
Hastings found
$|\la f_{j\sigma}^\dagger f_{i\sigma}\ra| = 2 \cdot 0.221$.
Using a reasonable $g_s = 4$, we then find $t = 0.66 J$.
On the other hand, from the study of the projected wavefunctions,
Ran\etal\cite{Ran06} suggest
\begin{equation}
t \approx 0.4 J \approx 70 K.
\label{tspinon}
\end{equation}
These are crude but reasonable estimates of the spinon band energy
scales.

\subsection{Pure System and System with a Single Impurity}
\begin{figure}[h]
\epsfxsize=3.5in \centerline{\epsfbox{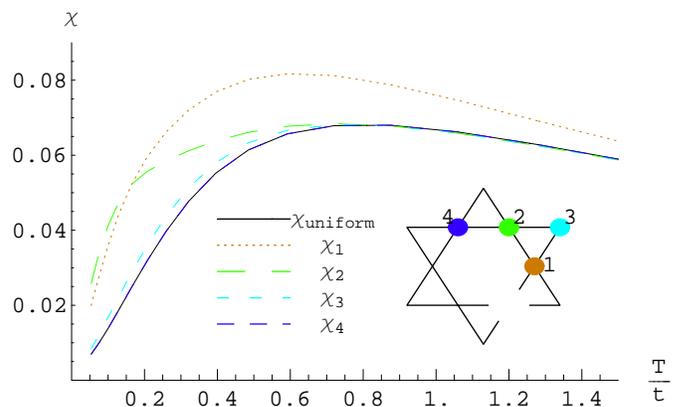}}
\caption{[Color Online]
The susceptibilities of first few neighbors near single impurity
in the spin liquid with $\pi$ fluxes through each hexagon on
Kagome lattice.  $\chi$ is in units $(g\mu_B)^2/t$.}
\label{fig_FF_SI}
\end{figure}

As mentioned, the local susceptibility near an impurity depends strongly
on how the spin liquid is modified near this impurity.
However, for illustration, we show the local susceptibility of the first
few neighbors for the case where we simply remove the appropriate links
from the fermion hopping problem.  For the size $44 \times  44 \times 3$
this is shown in Fig.~\ref{fig_FF_SI}.
The susceptibilities vanish linearly when $T \to 0$, since the pure
system is perturbed only by a single impurity here.
In our approximation, the susceptibility $\chi_1$ of the site next to
the impurity is larger than the bulk susceptibility.  This happens
because we take all bonds to have the same strength, while the
$\chi_1$ sites have fewer bonds.  This is an example where we believe
the simplified treatment near the impurity is inadequate and the
$\chi_1$ results should not be taken seriously.
A more accurate self-consistent treatment would likely require a
stronger hopping amplitude connecting two such sites next to an impurity,
and this would decrease $\chi_1$.
We discuss such possible improvements over the current treatment
in Sec.~\ref{sec:improvements}.

\begin{figure}[h]
\epsfxsize=3.5in \centerline{\epsfbox{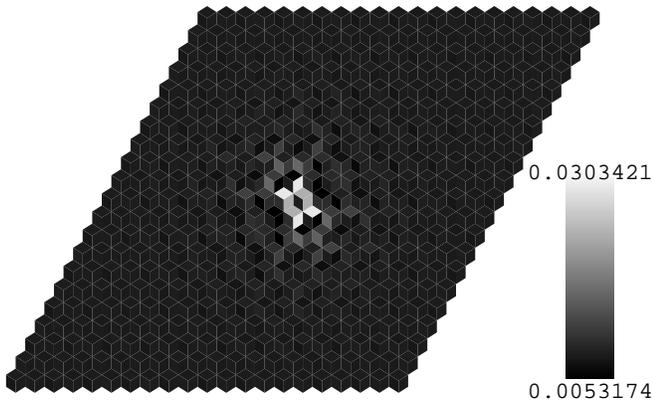}}
\caption{The susceptibilities near single impurity in the spin liquid
with $\pi$ fluxes through each hexagon on Kagome lattice at temperature
$T=0.054t$ where $t$ is the hopping amplitude.
In this figure, the color of a given point in the plane is given by the
value at the nearest Kagome site.}
\label{fig_FF_suscdensityplot}
\end{figure}

We also plot the shape of local susceptibility near a single impurity
in Figure~\ref{fig_FF_suscdensityplot}.
The ``ringing'' that we see is a more universal aspect of the Knight
shift texture around the impurity.  The corresponding wavevectors are
the mid-points of the Brilloin zone edges and are characteristic
for this spin liquid.  One can obtain some analytic understanding
of the mean field results by working perturbatively in the non-magnetic
impurity strength.  One finds that $(\chi_{\rm loc} - \bar\chi)/\bar\chi$
decays away from the impurity with a $1/r$ envelope on length scales
$1 \ll r \ll v/T$ and exponentially on larger scales
(here $v$ is the Fermi velocity, and we referenced $\chi_{\rm loc}$ by
$\bar\chi$ remembering that the susceptibilities vanish in the
$T \to 0$ limit).

\subsection{5\% Density of Impurities}
\begin{figure}[h]
\epsfxsize=3.5in \centerline{\epsfbox{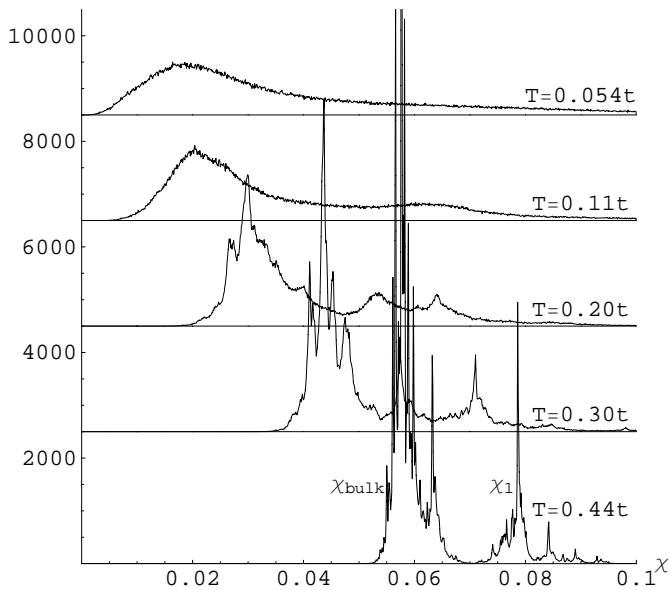}}
\caption{The histogram of susceptibilities on the Kagome lattice sites
in the spin liquid of fermionic spinons with $\pi$ fluxes through
hexagons in the presence of 5\% of impurities, treated by simply
removing the sites from the spinon hopping problem.
The histogram is the average of 300 samples of size
$20 \times 20 \times 3$.  $\chi$ is in units $(g\mu_B)^2/t$.
}
\label{fig_FF_susc}
\end{figure}

\begin{figure}[h]
\epsfxsize=3.5in \centerline{\epsfbox{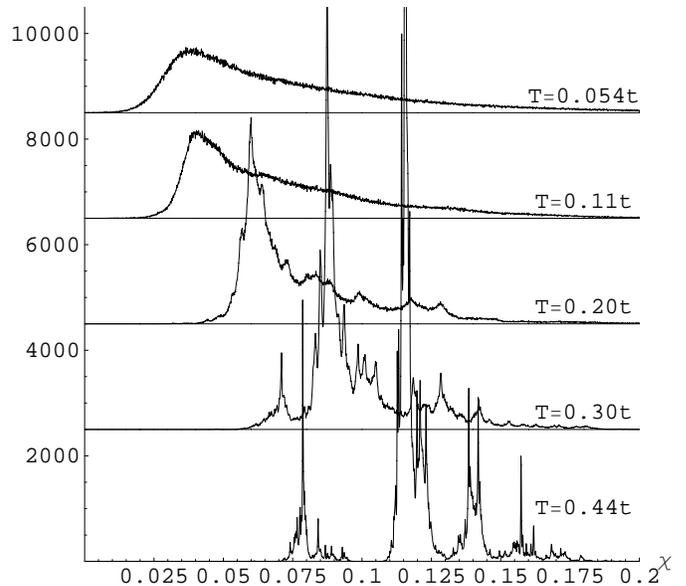}}
\caption{The histogram of susceptibilities on oxygen sites in the
spin liquid in the presence of 5\% of impurities.
The systems is the same as in Fig.~\ref{fig_FF_susc}.
The histogram is the average of 300 samples of size
$20 \times 20 \times 3$.}
\label{fig_FF_susc_pair}
\end{figure}

For the system with 5\% density of impurities, the histogram of
local susceptibilities at Kagome (Cu) sites is in
Figure \ref{fig_FF_susc} and at oxygen sites in
Figure \ref{fig_FF_susc_pair}.

We can understand the main features in the $\chi_{\rm loc}$
histograms as follows.  At the highest temperature shown, $T=0.44t$,
there are two main groups, just as in the high temperature series
analysis. The group with the higher $\chi$ comes from the sites that are
next to the impurities, while the other with the lower $\chi$
comes from the rest of the sites (bulk) and contains the main weight
of the sites.  As the temperature is lowered, the two features
move towards smaller susceptibilities, but the bulk one moves faster.
Also, between $T=0.4t$ and $T=0.2t$, we see a new feature
splitting away from the bulk and moving slower; we can roughly
identify this feature with the sites that have impurities
as their second neighbors.
The continuing marching of the bulk sites that are further from
the impurities to smaller $\chi$ is what the clean Kagome lattice
would do because of the vanishing density of states,
see Fig.~\ref{fig_FF_SI}.
As we lower the temperature, however, the influence of the impurities
spreads to further and further neighbors, and the local susceptibility at
the sites that already feel the impurities starts to slow down
and eventually saturates at a finite local $\chi$.

At $T=0.1t$, we already see no sharp features associated with specific
locations away from an impurity since all the sites already feel many
impurities.
The broad peak corresponds to the ``bulk'' sites that are furthest
from impurities; it remains visible since this group of sites contains
the largest volume fraction among the successive ``layers'' around
impurities.  The bulk peak stops its motion below $T=0.1t$;
its eventual location corresponds roughly to the typical impurity-induced
local density of states at these bulk sites.
Note that this value $\chi_{\rm peak} \approx 0.02/t$ is about two
times smaller than the average over all sites $\bar\chi = 0.0425/t$;
the latter contains contributions from the sites close to impurities
where the induced density of states is typically larger.

This behavior of the bulk peak is our main and robust prediction.
On the other hand, the behavior of the other peaks visible
at high temperatures depends on specific details of the treatment
of the impurity; these features should not be trusted in our
simplified approach and we will discuss the ways to improve the
treatment in Section~\ref{sec:improvements}.

Thinking about the NMR experiments, the bulk peak susceptibility
at low temperatures $\chi_{\rm peak} \approx 0.02/t \sim 0.04/J$,
where we used Eq.~(\ref{tspinon}).  This is about one half of the
bulk spin susceptibility at $T=2J$ from the high temperature
series study, Fig.~\ref{fig_local_susc_highT}.
In the $^{17}$O NMR experiments, this ratio is between one half
and one third, so our spin liquid results are reasonable.

To conclude this section, let us see if the disorder-induced
density of states can give sensible thermodynamic properties
in the Herbertsmithite.
In our treatment with 5\% impurity concentration, the DOS at the
Fermi level is
\begin{equation}
\nu(\epsilon_F) = 0.085/t
\end{equation}
per site per single spin species.
This would give average spin susceptibility of the Cu in the
Kagome layer
\begin{equation}
\bar{\chi} = 2 \nu(\epsilon_F) \mu_B^2 = 6.4 \cdot 10^{-4}
\times \left( \frac{100 \text{K}}{t} \right)
\quad \frac{\text{emu}}{\text{mol Cu}} ~.
\end{equation}
One should not take these numbers literally, since $\nu(\epsilon_F)$
depends quantitatively on the treatment near impurities, which
is very crude here.  Still, an estimate of $t$ like that in
Eq.~(\ref{tspinon}) produces $\bar\chi \sim 10^{-3}$ emu/mol~Cu
that compares reasonably with the experimental estimate of
the intrinsic Kagome layer susceptibility in Ref.~\onlinecite{Vries07}.
Furthermore, in the mean field treatment, spinons will contribute
to the specific heat as
\begin{equation}
c_v = 2\nu(\epsilon_F) \frac{\pi^2}{3} k_B^2 T  =
4.6 \times \frac{k_B T}{t}
\quad \frac{\text{Joule}}{\text{K} \cdot \text{mol Cu}} ~.
\end{equation}
which again yields reasonable numbers for temperatures of
several Kelvin.\cite{Helton06, Vries07}

\subsection{Gutzwiller Wavefunction Study of the Local Magnetizations}
In this section, we pursue the spin liquid proposal one step
beyond the mean field.
For the slave fermion approaches, one can achieve this by performing
the Gutzwiller projection of the singlet mean field state into the
space with precisely one fermion per site.  This approach is appealing
since one obtains a physical spin wavefunction, which can be viewed as a
variational state for an energetics study (and one can also study
properties of this spin state).
As we have already mentioned, Ran\etal\cite{Ran06} performed the
energetics study of the clean Kagome system and found the discussed
$\pi$-hexagon spin liquid to have the lowest trial energy.
We expect that the same construction on a lattice with vacancies will
have good energetics also for small concentration of impurities.
It is interesting to explore this issue in more detail in the future,
and we comment on this in Sec.~\ref{sec:improvements} below,
while here we want to take a crude look whether such proposal can
explain the observed inhomogeneous Knight shifts in the
Herbertsmithite.\cite{Imai07,Olariu07}
The specific results that we find here are as follows.  The spin
susceptibilities are always positive in the mean field.  We show that
they can become negative and the Knight shift distributions are
actually broader when we go beyond the mean field.

Unfortunately, one can not access the local susceptibilities from the
study of the singlet ground state.  To circumvent this, we consider trial
states with non-zero $S^z_{\rm tot}$ and study how the total spin is
distributed among the lattice sites.  The idea is that the pattern
of the local magnetizations $m^z_i = \la S^z_i \ra$ in such weakly
polarized states is similar to that of the local susceptibilities at
low temperatures.
This approach has been used in exact diagonalization studies of
spin systems with impurities, where one considers $m^z_i$ in the
lowest-energy states in the sectors with nonzero $S^z_{\rm tot}$.
We are trying this in the variational spin liquid studies in
the present Kagome problem and also for the organic spin liquid on the
triangular lattice.\cite{triangular_w_vacancy}
(We have also tried this approach in a one-dimensional chain with
vacancies\cite{1dchain} and reproduced the non-trivial qualitative
behavior predicted by Eggert and Affleck.\cite{Eggert95})

Consider a mean field state with different chemical potentials for the
two spin species, $\mu_\dn < \mu_\up$.  The orbitals with
$\mu_\dn < \epsilon < \mu_\up$ are occupied by the $\up$ spins but
not by the $\dn$ spins, which produces non-zero
$S^z_{\rm tot} = (N_\up - N_\dn)/2$.
In the mean field, the local magnetization is given by
\begin{equation}
m_{\rm loc}(i)
= \frac{1}{2} \sum_{\mu_\dn < \epsilon_n < \mu_\up}
|\psi_n(i)|^2 ~.
\label{eq:mloc}
\end{equation}
Observe the similarity of this expression to that for the local
susceptibility at finite temperature, Eq.~(\ref{eq_chi_freeF}).
The difference is that here the contributing orbitals are selected by
a sharp energy window $[\mu_\dn, \mu_\up]$,
while in Eq.~(\ref{eq_chi_freeF}) the window is soft proportional to
$(-\partial f/\partial\epsilon)$ with width set by the temperature.
If there is a finite density of states at the Fermi level and if the
properties of the orbitals do not change strongly with the energy,
the two windows of similar width should give similar results for the
local magnetization or susceptibility normalized by the corresponding
average values.  The above two windows roughly match when
$\mu_\up - \mu_\dn \approx 4T$.  In our numerical calculations,
we indeed observe such correspondence of the $m_{\rm loc}$ and
$\chi_{\rm loc}$ distributions.

\begin{figure}
\centerline{\includegraphics[width=\columnwidth]{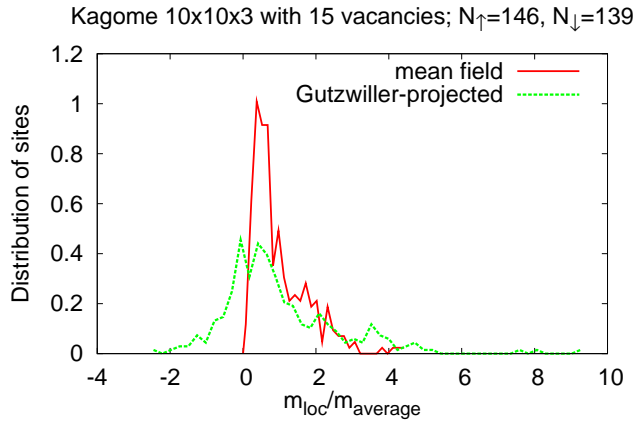}}
\vskip -2mm
\caption{[Color Online] Distribution of the local magnetization in the
$\pi$-flux spin liquid state with non-zero $S^z_{\rm tot}$.
The sample is $10 \times 10 \times 3$ with $15$ randomly placed
vacancies;
the total magnetization is $S^z_{\rm tot} = (N_\up - N_\dn)/2 = 7/2$.
We compare the mean field and the Gutzwiller-projected results for
exactly the same system and find that the projection broadens the
histogram by about a factor of 1.5-2, which can be clearly seen despite
the noisiness of the data in the single random sample.
}
\label{fig:distrSzloc}
\end{figure}

From the mean field state with non-zero $S^z_{\rm tot}$, we construct
the spin wave function by Gutzwiller projection.  Note that
$S^z_{\rm tot}$ is preserved by the projection, so the average
magnetization is unchanged.
However, the local magnetizations are modified.
Our observations in the present system can be summarized by the following
crude rule:  The deviation of the site magnetization from the average
is enhanced by the projection by a typical factor between 1.5 and 2.
This implies that the distribution of the local magnetization
is broadened by the same factor upon the projection.
Figure~\ref{fig:distrSzloc} illustrates this for a
$10 \times 10 \times 3$ sample with $15$ vacancies and
$N_\up - N_\dn = 7$.

We have tried several lattice sizes, different disorder samples,
and different magnetizations $S^z_{\rm tot} = (N_\up - N_\dn)/2$,
and in all cases observed the above rule.
This rule is not strict but holds for a majority of sites.
More precisely, the histogram of the ratios
$(m_{\rm loc}^{\rm Gutzw} - \bar{m})
  / (m_{\rm loc}^{\rm meanfield} - \bar{m})$
is peaked near $1.5 - 2$, which is the claimed typical enhancement
factor for the deviations from the average.
In particular, while the mean field local magnetization is always
non-negative, see Eq.~(\ref{eq:mloc}), the local magnetization can
become negative after the projection.  It can be opposite to the
bulk magnetization because of the antiferromagnetic correlations
present in the system.

Let us say few words about our choice of $N_\up, N_\dn$ in
Fig.~\ref{fig:distrSzloc}.
On one hand, if we take $N_\up - N_\dn = 1$, we worry if the results
are dominated by the specific single-particle state that happened to lie
at the Fermi level.  This is a concern because of our view that all
states will be eventually Anderson-localized.
However, by considering the participation ratio for the states near
the Fermi level in the present sample, we find that they are still
spread over about one third of the sites.
We have checked that even for such small $N_\up - N_\dn$ the
results for the broadening of the distribution by the projection
are qualitatively similar.
On the other hand, we do not want $N_\up - N_\dn$ to be large,
which could take us far from the region in the spectrum near the
Fermi level and also would correspond to physically inaccessible
magnetization.  Our $N_\up=146, N_\dn = 139$ is a compromise,
since a few typical states around the Fermi level are sampled,
and also since it corresponds to about $2.5$\% of the maximal
polarization, which is reasonable also from the experimental
perspective.  For example, using the estimate\cite{Bert07}
$\chi_{\rm intrinsic} = 1.25 \cdot 10^{-3}$ emu/mol Cu,
the field of 7 Tesla in the $^{17}$O experiments\cite{Olariu07}
magnetizes the system to about 1.5\% of the maximal
polarization of the Kagome layer.

\begin{figure}
\centerline{\includegraphics[width=\columnwidth]{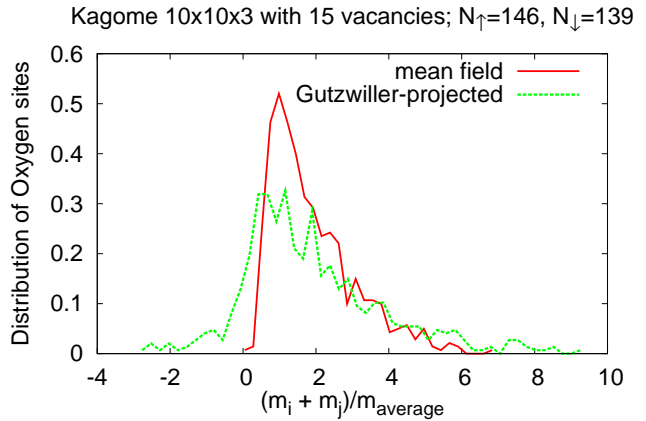}}
\vskip -2mm
\caption{[Color Online]
Oxygen NMR local fields corresponding to the same system as
in Fig.~\ref{fig:distrSzloc}.
}
\label{fig:distrOxygenNMR}
\end{figure}

To conclude this section, we find that the Gutzwiller trial
wavefunction treatment does not modify the mean field results
qualitatively but widens the magnetization distributions by about a
factor of two.\cite{renorm4mloc}
In particular, at low temperature the distribution will have a
sizable wing on the negative values, which is consistent with
the NMR experiments.\cite{Imai07, Olariu07}
To describe the $^{17}$O NMR line, we need to histogram the
field $m_i + m_j$ associated with bonds $<ij>$ of the Kagome lattice.
This is shown in Fig.~\ref{fig:distrOxygenNMR}.  The wing at negative
fields is weaker than in the Cu NMR because the oxygens perform some
``averaging'' of the local $m_i$ (while the total magnetization
is positive).
We reiterate here the crudeness of our study;
in particular, we are not attempting here to explain the actual
experimental histogram in the $^{17}$O NMR, which looks nearly
symmetric with respect to the peak at the lowest temperature.
Our main message is that within the spin liquid approach, by
going beyond the mean field, we expect broader distribution
of the Knight shifts with both positive and negative
susceptibilities due to the underlying correlations.

\subsection{Issues with the Spin Liquid Treatment and Possible
Improvements}
\label{sec:improvements}
The main reservation about the presented spin liquid study is that
we modeled the effect of impurities by simply removing the Zn sites
in the spinon mean field.
We did not perform a self-consistent mean field or an energetics
study of the system with impurities.
However, we did think how such more accurate treatments could
affect our results and tested the robustness of the features
by trying different modifications that such treatments might
require.

One minor concern is that even though the system is inhomogeneous,
we require the half-filling for spinons only on average over the
whole sample and not at each site.  The latter is likely a better
treatment since then the Gutzwiller projection is less drastic
and the mean field is closer to the projected state.
We find that in our samples the deviations of the local densities
from the average are small and smaller in relative terms than the
deviations of the local susceptibilities that we are interested in.
We also tested small systems achieving the half-filling precisely
at all sites by adjusting local chemical potential and found that
the susceptibilities did not change qualitatively or significantly
quantitatively and only at the the lowest temperatures the
histograms became somewhat broader.

A more serious concern is that we took the spinon hopping amplitudes
to be the same as in the clean system, instead of finding them
self-consistently.
Thus, we expect that at low temperatures the hopping connecting
two sites that are both next to an impurity should be larger than
for other bonds.  Consider, for example, the renormalized mean field
scheme, Eq.~(\ref{RMF}).
When we calculate the bond expectation value
$\zeta_{ij} = \la f_j^\dagger f_i \ra$
in our mean field starting with all $t_{ij}$ equal,
which can be viewed as a first step in the self-consistency iteration,
we find that it is the largest on the bonds next to impurities.
We have not pursued this scheme any further since in the present case
it would actually lead to a dimer state in the clean system,
which we know is a pathology of such treatment when applied to the
spin-1/2 system.
Roughly, one should not be using the same renormalization factor $g_s$
irrespective of the bond expectation value.  By comparing with the
Gutzwiller-projected measurements, one typically uses $g_s = 4$ for
translationally invariant states, but one should take a smaller
$g_s = 2$ for the dimer states.
One could use a different mean field scheme to capture this,
e.g., the scheme due to Hsu,\cite{Hsu90, Hastings01}
which leads to the self-consistency conditions of the form
\begin{equation}
t_{ij} = \frac{3 J_{ij} \zeta_{ij} (1-16|\zeta_{ij}|^4)}
              {(1 + 16 |\zeta_{ij}|^4)^2} ~.
\end{equation}
This suppresses the dimerization tendency of the less elaborate
mean field but may be doing it already too much, while we may want
to retain some possibility of local dimers in the system with impurities.
We leave a detailed exploration of such approaches and whether
they can capture the actual energetics in the system with impurities to
future work.

It is interesting to notice however that if we took the most naive
treatment Fig.~\ref{fig_FF_SI}, at low enough temperatures the nearest
neighbor susceptibility is larger than twice the bulk and a little peak
starts to emerge on the right of the bulk peak in
Fig.~\ref{fig_FF_susc_pair} for $T \approx 0.11t$.
If this were the case then the peaks in Ref.~\onlinecite{Olariu07}
at low temperatures would have the opposite interpretation -- bulk peak
would become closer to zero and the next to impurity oxygen peak further
from zero.  However we cannot make this conclusion at this point since
the bonding of the spins next to the impurity will decrease the local
susceptibility near the impurity.  We only would like to point out that
one should be careful when interpreting the experiment.

For the treatment of the spin liquid near impurity, we only note some
crude things that we tried to see how our results might be affected.
Motivated by the above observation,
we have studied local susceptibilities at low temperatures when
we multiplied by 2 all the bonds connecting the two sites that are both
next to an impurity.  Certainly, this has a strong effect on the
susceptibility $\chi_1$ of these sites -- it in fact becomes smaller
than the bulk one, which is opposite to the results in the preceding
section where the $\chi_1$ is the largest.  Thus, such treatment can
make the results for $\chi_1$ more in line with the series study,
Sec.~\ref{sec:series}, and exact diagonalization study,
Ref.~\onlinecite{Dommange03}, that suggest that this susceptibility is
depressed at low temperatures.  On the other hand, the above change
of the hopping strengths near impurities does not have significant
effect on the bulk susceptibility histogram peak and its marching to
lower values with decreasing temperature (except, of course, for
small numerical differences).
We have also tried modifying not only the strengths of the bonds
but also fluxes next to impurities.  For example, if we view the
proposed $\pi$-hexagon spin liquid as a time reversal invariant
way of performing flux attachment and flux-smearing mean field,
obtaining Dirac spin liquid instead of a chiral spin liquid
(see discussion in Sec.~IIC of Ref.~\onlinecite{DBL}), it suggests that
we should also remove a $\pi$ flux when we remove a spin.
In an exploratory Gutzwiller energetics study with impurities,
we indeed find that this often improves the trial energy.
As far as we are concerned here with the local susceptibilities,
such local flux modification of course has an effect on the sites
near the impurities, but has little effect on the described
qualitative behavior of the bulk peak.

We also remark that making all hexagon fluxes equal to zero\cite{Ma08}
results in a qualitatively different behavior of the bulk of the
susceptibility histogram.  Here the clean system has a finite
density of states at the Fermi level and therefore a nonzero
susceptibility at low temperatures.
We tried this for the system with 5\% density of impurities and
indeed found that the bulk peak doesn't march down.  Instead,
it approaches a relatively large value of susceptibility and broadens
to both high and low values of susceptibility.  We then propose that
the experiments suggest more a Dirac-like spin liquid than the one
with a full Fermi surface.


\section{Conclusions}
Using the high temperature series expansion, we calculated local
susceptibility for the nearest neighbor Heisenberg antiferromagnet on
Kagome lattice down to about $T\approx 0.4J$.  The resulting histogram
of susceptibilities can be directly compared to the experiment to test
whether the Heisenberg model with impurities applies to
ZnCu$_3$(OH)$_6$Cl$_2$.  The calculated histogram contains a lot of fine
structure some of which we hope can be observed if the resolution of the
experiments improves.
One issue not considered in experiments is that the local susceptibility
$\chi_1$ next to impurities is larger than the bulk value by as much as
20\% at high temperatures between $0.5 J$ and $2J$.  This should affect
the peak associated with the oxygens close to impurity, and one should
be careful at interpreting the observed features at the intermediate
and lower temperatures.
Our work suggests that $\chi_1$ drops below the bulk value at low
temperatures and is consistent with earlier exact diagonalization
study\cite{Dommange03} which found that these sites tend to form
singlets.  The latter was invoked in Ref.~\onlinecite{Olariu07}
to explain a sharp feature in the NMR signal appearing near low
temperatures.  This is appealing, but more work needs to be done to
understand the origin of this feature, particularly in the presence
of the sizable impurity concentration, and also with the added
bulk peak contribution that we expect in the spin liquid regime.

At low temperatures we assumed that the system forms a spin liquid of
Ran\etal\cite{Ran06}  We calculated the susceptibility
histograms in the mean field theory and went beyond mean field by using
Gutzwiller projection.  Our results for the local susceptibility should
not be trusted quantitatively close to impurity, since these depend on
the details of the treatment of the spin liquid near the impurities,
while we took a crude approach and only outlined how one can make it
more accurate.  In particular, we are not able to discuss histogram
features associated with sites close to impurities.
However we have a robust prediction for the behavior of the peak
coming from the bulk of the sites (Fig.~\ref{fig_FF_susc}):
The bulk peak marches down with decreasing temperature; its weight is
gradually spread as further layers of sites around the impurities
start to feel their share of the impurity-induced density of states;
the broadening saturates when all sites start to feel several and more
impurities.  We expect this qualitative description to be insensitive
to the details of the mean field in the immediate vicinity of the
impurities.  It is sufficient to know that the clean spin liquid has
Dirac density of states and that the bulk sites do not see the
impurities down to some temperature scale, while any treatment near the
impurities represents some local disorder that always induces some
density of states at the bulk sites.  The effect of the Gutzwiller
projection is crudely to increase the variation of the local
susceptibility by about a factor of two.  In particular, as the peak
spreads some of the susceptibilities become negative.

This behavior is consistent with the observed Knight shifts in the
ZnCu$_3$(OH)$_6$Cl$_2$ experiments.  Such behavior observed in the
presence of a sizable density of impurities then suggests that the
clean spin liquid would have a vanishing density of states at the
Fermi level.
Thus it distinguishes this spin liquid from the proposal with a spinon
Fermi surface for which the peak would not march down but instead
saturate at a large positive value.
On the other hand, since the Algebraic Vortex Liquid of
Ryu\etal\cite{AVL} has Dirac-like gapless character, it would likely
produce inhomogeneous Knight shift behavior similar to the Dirac
spin liquid.

In the future work it would be good to treat the spin liquid right
near impurities more accurately, e.g., in a self-consistent mean field
or in an energetics study, and consequently make a reliable prediction
for the spin susceptibility at these sites.
Quantitative comparison with other approaches such as exact
diagonalization would also be useful.
It would be interesting to consider quantitatively the gauge
theory description of such spin liquid with disorder-induced
finite density of states, though we do not expect our mean field
results to be modified significantly.
In a companion work,\cite{triangular_w_vacancy} we are studying
non-magnetic impurities in the triangular lattice antiferromagnet and
explore spin liquid with a spinon Fermi surface, which may be relevant
for the organic spin liquid material $\kappa$-(ET)$_2$Cu$_2$(CN)$_3$.


\acknowledgments
We thank J.~Alicea, M.~P.~A.~Fisher, S.~Ryu, and R.~R.~P.~Singh
for useful discussions and the A.~P.~Sloan Foundation for financial
support (OIM).



\begin{thebibliography}{99}
\bibitem{1dchain}
In a one-dimensional (1d) chain with defects or with open boundaries,
we studied the magnetization distribution in a projected Fermi sea state
with small non-zero $S^z_{\rm tot}$.
In the mean field, $m(x)$ oscilates along the chain as $1 + (-1)^x$
with a constant amplitude.  We find that after the projection the
staggered piece in fact grows away from the impurity as
$\sqrt{x} (-1)^x$.
This is consistent with the Eggert and Affleck prediction for the
local susceptibility in the $T=0$ limit,\cite{Eggert95}
which is responsible for very strong NMR line broadening
observed in 1d spin chain materials with impurities.
On the other hand, in all our studies in two dimensions, we
only found a fixed numerical enhancement rather than change
in the qualitative behavior after the projection.
\bibitem{renorm4mloc}
The broadening of the histograms of $m_{\rm loc}$ by about a factor
of two upon the projection can be understood in the spirit of
``Gutzwiller renormalizations''.
For example, if we take a partially polarized mean field state and
perform the projection exactly at one site, the corresponding local
magnetization is increased by a factor
$1/(\la n_\up \ra + \la n_\dn \ra - \la n_\up \ra \la n_\dn \ra)$,
where the expectation values are in the free fermion system.
For weak polarization $\la n_\up \ra \approx \la n_\dn \ra \approx 1/2$,
this factor is roughly 2.  This simple reasoning treats only one site
and does not take into account additional inter-site correlations
introduced by the projection.  This seems to work fine in the
two-dimensional spin liquids that we studied but fails in one
dimension.\cite{1dchain}
\end{thebibliography}
\end{document}